\documentclass[twocolumn,superscriptaddress]{revtex4-2}

\usepackage[usenames,dvipsnames]{xcolor}

\usepackage{lettrine}

\usepackage{amsfonts, amsmath, amssymb}
\usepackage[utf8]{inputenc}
\usepackage[english]{babel}
\usepackage{graphicx}
\usepackage{braket}
\usepackage{wrapfig}
\usepackage{hyperref}
\usepackage{bm}
\usepackage{color}
\usepackage{soul}
\usepackage{mathtools}
\usepackage{float}
\usepackage{ulem}
\usepackage{cancel}

  {\left\lbrace\begin{array}{@{}l@{}}}%
  {\end{array}\right.}

\newcommand{\blank}{\vspace{3mm}\noindent}

\begin{document}

\title{Controlling topological phases of matter with quantum light}

\author{Olesia Dmytruk}
\email{olesia.dmytruk@college-de-france.fr}
\affiliation{JEIP, USR 3573 CNRS, Coll\`{e}ge de France, PSL Research University, 11 Place Marcelin Berthelot, 75321 Paris Cedex 05, France}

\author{Marco Schir\`o}
\email{marco.schiro@college-de-france.fr}\thanks{On Leave from: Institut de Physique Th\'{e}orique, Universit\'{e} Paris Saclay, CNRS, CEA, F-91191 Gif-sur-Yvette, France}
\affiliation{JEIP, USR 3573 CNRS, Coll\`{e}ge de France, PSL Research University, 11 Place Marcelin Berthelot, 75321 Paris Cedex 05, France}

\begin{abstract}
Controlling the topological properties of quantum matter is a major goal of condensed matter physics. A major effort in this direction has been devoted to using classical light in the form of Floquet drives to manipulate and induce states with non-trivial topology. A different route can be achieved with cavity photons. Here we consider a prototypical model for topological phase transition, the one-dimensional  Su-Schrieffer-Heeger (SSH) model, coupled to a single mode cavity. We show that quantum light can affect the topological properties of the system, including the finite-length energy spectrum hosting edge modes and the topological phase diagram. In particular we show that depending on the lattice geometry and the strength of light-matter coupling one can either turn a trivial phase into a topological one or viceversa using quantum cavity fields. Furthermore, we compute the polariton spectrum of the coupled electron-photon system, and we note that the lower polariton branch disappears at the topological transition point. This phenomenon can be used to probe the phase transition in the SSH model.
\clearpage
\end{abstract}
\maketitle

\lettrine{L}{ight}-control of quantum materials is an emerging goal of condensed matter physics. While traditionally light has played the role of spectroscopic probe, recent experimental efforts have demonstrated the possibility of inducing novel phases of matter by selective irradiation~\cite{basov2017towards,disa2021engineering} or by coupling to cavity fields~\cite{schlawin2021cavity,garciavidal2021manipulating,valmorra2021vacuum}.

Topological phases of matter~\cite{hasan2010colloquium,qi2011topological} play an important role in this perspective due to their robustness and their possible application in quantum technologies. As such there has been tremendous interest in the possibility of enhancing or inducing topological properties in electronic system by light, a notable example being the Floquet topological insulator~\cite{oka2009photovoltaic,lindner2011floquet}. 
Floquet-engineered topological band structure was experimentally demonstrated in a variety of solid state materials irradiated by circularly polarized light~\cite{wang2013observation,mciver2020light}. Similar efforts in realizing topological phases of matter using time-modulated optical lattices have been made with ultracold atoms~\cite{cooper2019topological}. Quantum fluctuations of the light field in a cavity offer new possibility to  probe, control and tune the properties of a material, leading for example to polaritons, hybrid light-matter excitations, with non-trivial topological properties~\cite{karzig2015topological,ohm2015microwave,trif2012resonantly,contamin2021hybrid} or to anomalous Hall response in presence of a circularly polarised field~\cite{wang2019cavity}.  Here we ask the question of what happens to a topological material due to the coupling to cavity photons. Such question is of direct experimental relevance given the recent observation of a breakdown of topological protection in an integer Quantum Hall system coupled to a cavity~\cite{appugliese2022breakdown} and can be also explored with ultracold atoms embedded in high-finesse resonators~\cite{roux2020strongly,mivehvar2017superradiant}.

A prototypical model for topological behavior in one dimension is the Su-Schrieffer-Heeger (SSH) model~\cite{su1979solitons}, originally introduced to describe conducting electrons in polyacetylene and which later found applications in a variety of settings~\cite{delplace2011zak,gomez2013floquet,dallago2015floquet,goren2018topological,downing2019topological,nie2021dissipative,perez2021topology}. The SSH model describes  a 1D dimerized chain with alternating weak and strong nearest-neighbor hoppings  and displays a topological transition associated with a non-trivial Zak phase. The experimental realization of the SSH model has been demonstrated  in ultracold atoms~\cite{atala2013direct,meier2016observation,deleseleuc2019observation}, graphene nanoribbons~\cite{rizzo2018topological,groning2018engineering} and in various platforms for topological phenomena, from photonics~\cite{ozawa2019topological,Solnyshkov:21,kim2021quantum} to mechanical metamaterials~\cite{huber2016topological}.

In this work, we study the interplay of topological phases of matter and quantum light by considering the SSH model coupled to a single mode photonic resonator through the full gauge invariant Peierls phase. We show that quantum fluctuations of the light field can have dramatic effects on the topological properties of the system and most remarkably turn a trivial insulating phase into a topological one above a critical light-matter coupling, with emergence of non-trivial topological edge modes.  This topological transition is possible due to the structure of the Peierls phase, which preserves chiral symmetry and introduces a non-trivial dependence from the lattice geometry, similarly to the case of classical light explored in the context of Floquet driving. Differently from the latter, however, our results do not suffer from heating runaway problems that often plagued Floquet engineering schemes and required high-frequency driving or tailored dissipation. Finally, when considering the limit of an infinite chain we show that the topological transition reduces to the case of the isolated SSH model. However, photons retain memory of the non-trivial topology of the system as we show by computing the polariton spectrum of the hybrid system. We find that the lower polariton branch disappears when the hopping amplitudes is tuned to satisfy the topological criterion. Thus, measurements of the polariton frequencies could provide direct access to the topological phase transition point.

\begin{figure}[t!]
	\includegraphics[width=.9\columnwidth]{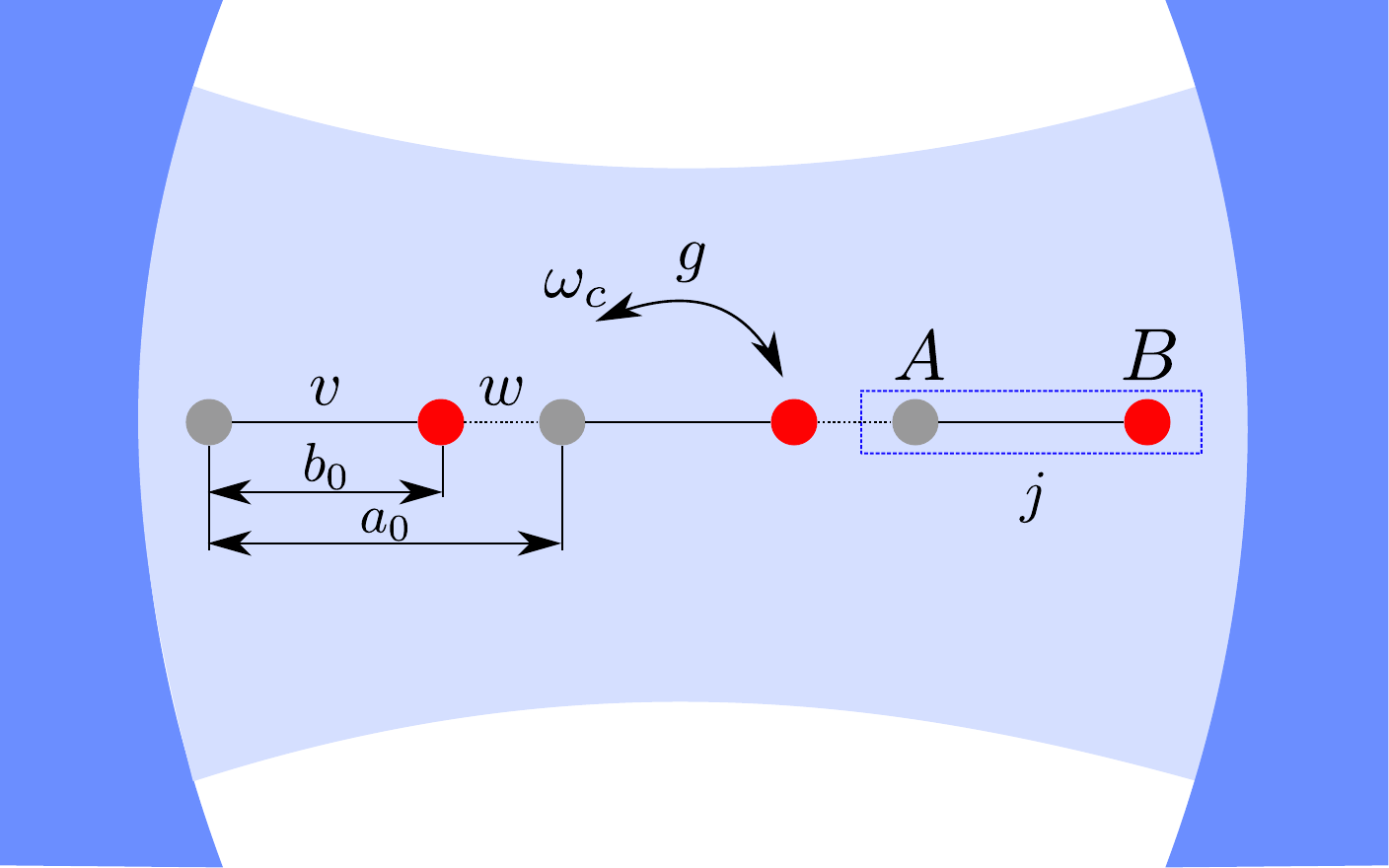}
	\caption{ Scheme of the setup: A one-dimensional dimerized SSH chain with intracell $v$ (black solid line) and intercell $w$ (black dashed line) hopping amplitudes coupled to a single mode cavity with frequency $\omega_c$. The strength of the light-matter coupling is given by $g$. The lattice constant is $a_0$, and
		the distance between the sublattices $A$ and $B$ within the same unit cell (blue dashed square) is $b_0$. 
	}
	\label{fig:Scheme}
\end{figure}

\blank
\textbf{Results }\\
{\bf The Model.}
We consider a one-dimensional SSH model coupled to a single mode cavity (see Fig.~\ref{fig:Scheme}). 
The SSH model describes spinless electrons hopping on a one-dimensional chain composed of $L$ unit cells with two sublattices $A$ and $B$.
The Hamiltonian reads
\begin{align}
H_{\rm SSH} &= v\sum_{j=1}^{L}\left( c^\dag_{jA}c_{jB}+c^\dag_{jB}c_{jA}\right) \notag\\
&- w\sum_{j=1}^{L-1}\left(c^\dag_{j+1,A}c_{jB}+c^\dag_{jB}c_{j+1,A}\right),
\label{eq:HamiltonianSSH}
\end{align}
where $v\geq 0$ is the intracell hopping amplitude, $w \geq 0$ is intercell hopping amplitude, and $c^\dag_{j\mu}$ ($c_{j\mu}$) are the fermionic creation (annihilation) operators at site $j$ and sublattice $\mu = A,B$.  In absence of any light-matter interaction the SSH model displays a well-known topological phase transition as the ratio between the two hoppings is tuned. In particular the ground-state of the system is in  topological phase for $v < w$ with finite Chern number and in a topologically trivial phase for $v>w$. We emphasize that in this case the spectrum does not depend on the lattice spacing $b_0$.
The single mode cavity Hamiltonian is given by 
\begin{align}
H_{\rm ph} = \omega_c \left(a^\dag a + \dfrac{1}{2}\right),
\label{eq:Hphoton}
\end{align}
where $\omega_c$ is the mode frequency and $a^\dag$ ($a$) are the photon creation (annihilation) operators satisfying $[a, a^\dag] = 1$.  We couple the SSH chain to the cavity mode through the gauge invariant Peierls substitution~\cite{li2020electromagnetic,sentef2020quantum,Li2020manipulating,guerci2020superradiant,dmytruk2021gauge}, which amounts to dress the hoppings amplitudes entering the SSH Hamiltonian Eq.~\ref{eq:HamiltonianSSH} as $v\rightarrow v\,e^{ie\mathcal{A}\ell_{AB}}$ and $w\rightarrow w e^{-ie\mathcal{A}\ell_{BA}}$, where  $e$ is the electric charge, $\mathcal{A}=A_0(a+a^{\dagger})$ is the uniform vector potential, and 
$\ell_{AB}=b_0,\ell_{BA}=a_0-b_0$ represent respectively the distance between atoms in the same (different) unit cell. To understand the origin of this different renormalization we note that the Peierls substitution is completely equivalent to applying a unitary transformation $\Omega$ to electronic Hamiltonian~\cite{dmytruk2021gauge}, i.e.
\begin{align}
H= H_{\rm ph} + \Omega^\dag H_{\rm SSH}\Omega,
\end{align}
where the unitary is defined as
\begin{align}\label{eq:UP}
\Omega  = e^{ie\mathcal{A}\sum_{j\mu}R_{j\mu} c^\dag_{j\mu}c_{j\mu}}.
\end{align}
Here, $R_{j\mu}$ indicates the position of the atom in the $\mu=A,B$ sublattice, with $R_{jA}=ja_0$ and $R_{jB}=ja_0+b_0$. Under this unitary transformation the fermionic operators $c_{jA},c_{jB}$ acquire a site-dependent phase (See Methods) which leads to the full light-matter Hamiltonian of the form
(setting $a_0=1$)
\begin{align}
	&H=  v\sum_{j=1}^{L}\left(e^{i \frac{g}{\sqrt{L}} b_0 (a+a^\dag)}c^\dag_{jA}c_{jB}+\text{h.c.}\right)\notag\\
	 &- w\sum_{j=1}^{L-1}\left(e^{-i \frac{g}{\sqrt{L}} (1-b_0) (a+a^\dag)}c^\dag_{j+1,A}c_{jB}+\text{h.c.}\right)\notag\\
	 &+  \omega_c \left(a^\dag a + \dfrac{1}{2}\right).
	 \label{eq:PeierlsHamiltonian}
\end{align}
Several observations are in order here concerning this Hamiltonian. First, the two hopping integrals $v,w$ are renormalised by different Peierls factors due to the different distance between atoms within the same or different unit cell. This gives an explicit dependence of the Hamiltonian from the ratio of lattice spacing $b_0$, which instead is missing in equilibrium and will play a key role in the following.   Furthermore, we note that using the unitary operator defined in Eq.~(\ref{eq:UP}) one can write down a fully equivalent light-matter Hamiltonian in the Dipole Gauge (see Methods). Finally, we note that in Eq.~(\ref{eq:UP}-\ref{eq:PeierlsHamiltonian}) we have rescaled light-matter coupling $g=eA_0$ by the factor $1/\sqrt{L}$, namely, we consider the so-called collective ultra-strong coupling regime~\cite{ciuti2005quantum,frisk2019ultrastrong,pilar2020thermodynamics,eckhardt2021quantum}. While this ensures that in the thermodynamic limit a single cavity mode does not affect thermodynamic bulk properties, as we indeed are going to recover, we show here that the topological properties, in particular the emergence of edge modes in the finite-length energy spectrum is strongly modified by the coupling to light.

{\bf Energy spectrum of finite-length SSH model coupled to photons}. A signature of the non-trivial topology of the SSH model is contained in its finite-length energy spectrum which hosts edge modes exponentially localised near the boundaries~\cite{asboth2016short}. The first question we address is therefore what is the fate of the edge modes in presence of a finite coupling to the cavity mode. 
To this extent we solve the model in mean field (see Methods) that corresponds to neglecting correlations between the cavity modes and electrons,  which is justified in the limit of large $L$. Within this ansatz we obtain a renormalised SSH model with hopping amplitudes $v$ and $w$ dressed by the cavity photon
\begin{align}
&H_{\rm el}^{\rm mf}= \langle \phi | H | \phi \rangle = 
\sum_{j=1}^{L} \left( \tilde{v}\, c^\dag_{jA}c_{jB}+\text{h.c.} \right)\notag\\
&- \sum_{j=1}^{L-1}   \left( \tilde{w}\,c^\dag_{j+1,A}c_{jB} + \text{h.c.}\right),
\label{eq:HelMF}
\end{align}
where we introduced the renormalized hoppings $\tilde{v}=v\varphi(g,b_0),\tilde{w}=w\varphi(g,1-b_0)$ with the renormalization factor 
	\begin{align}
	\label{eq:Ema0b0}
\varphi(g,\ell)= \langle\phi|e^{-i \frac{g \ell}{\sqrt{L}}(a+a^\dag)}|\phi\rangle.
	\end{align}

\begin{figure}[t!]
	\centering
	\includegraphics[width=0.9\linewidth]{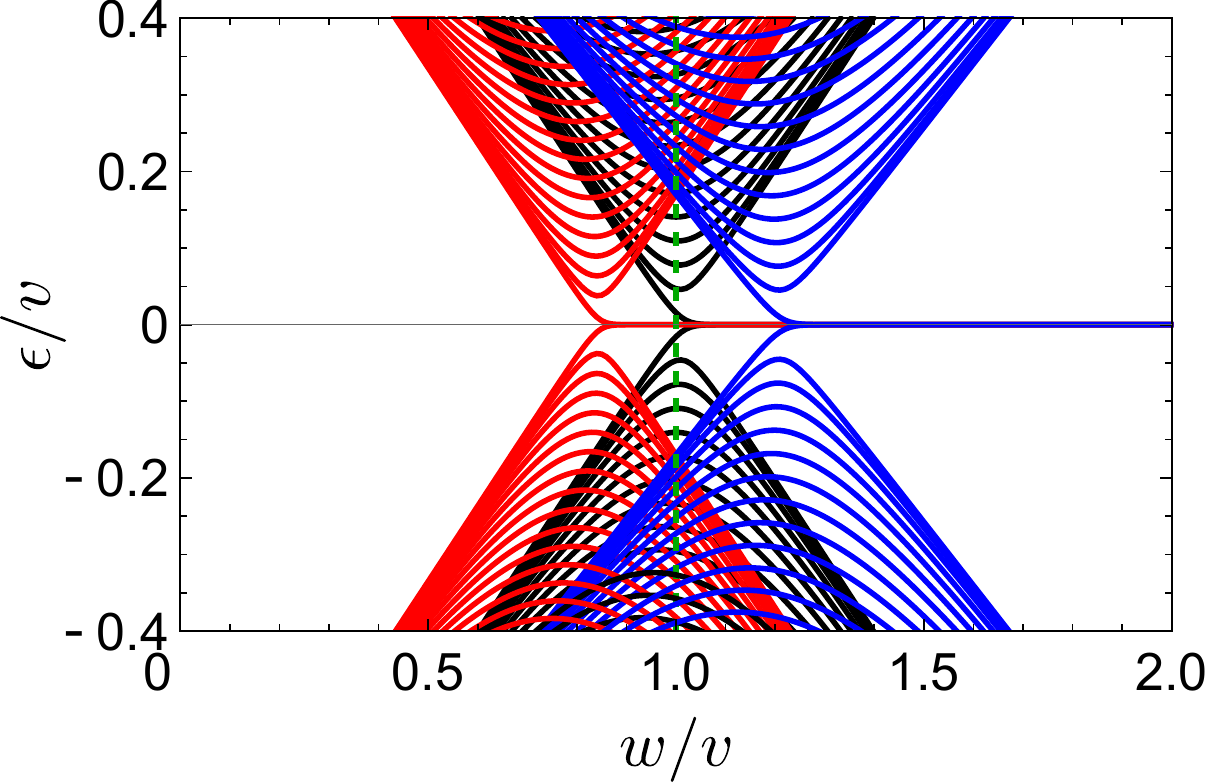}
	\caption{Energy spectrum of the SSH model coupled to a single mode cavity as a function of $w/v$. Black solid lines correspond to $g = 0$ (no coupling to the cavity) and red (blue) solid lines correspond to $b_0 = 0.8$ ($b_0 = 0.2$) and $g = 20$. Green dashed line corresponds to the topological phase transition point for $g = 0$, namely $w/v=1$.  Other parameters are chosen as $v = 1$, $L = 100$, $a_0 = 1$, $\omega_c=0.5$, and maximum number of photons $N_{max} = 10$.
	}
	\label{fig:SpectrumFinite}
\end{figure}

\begin{figure*}[t!]
	\centering
	\includegraphics[width=0.9\linewidth]{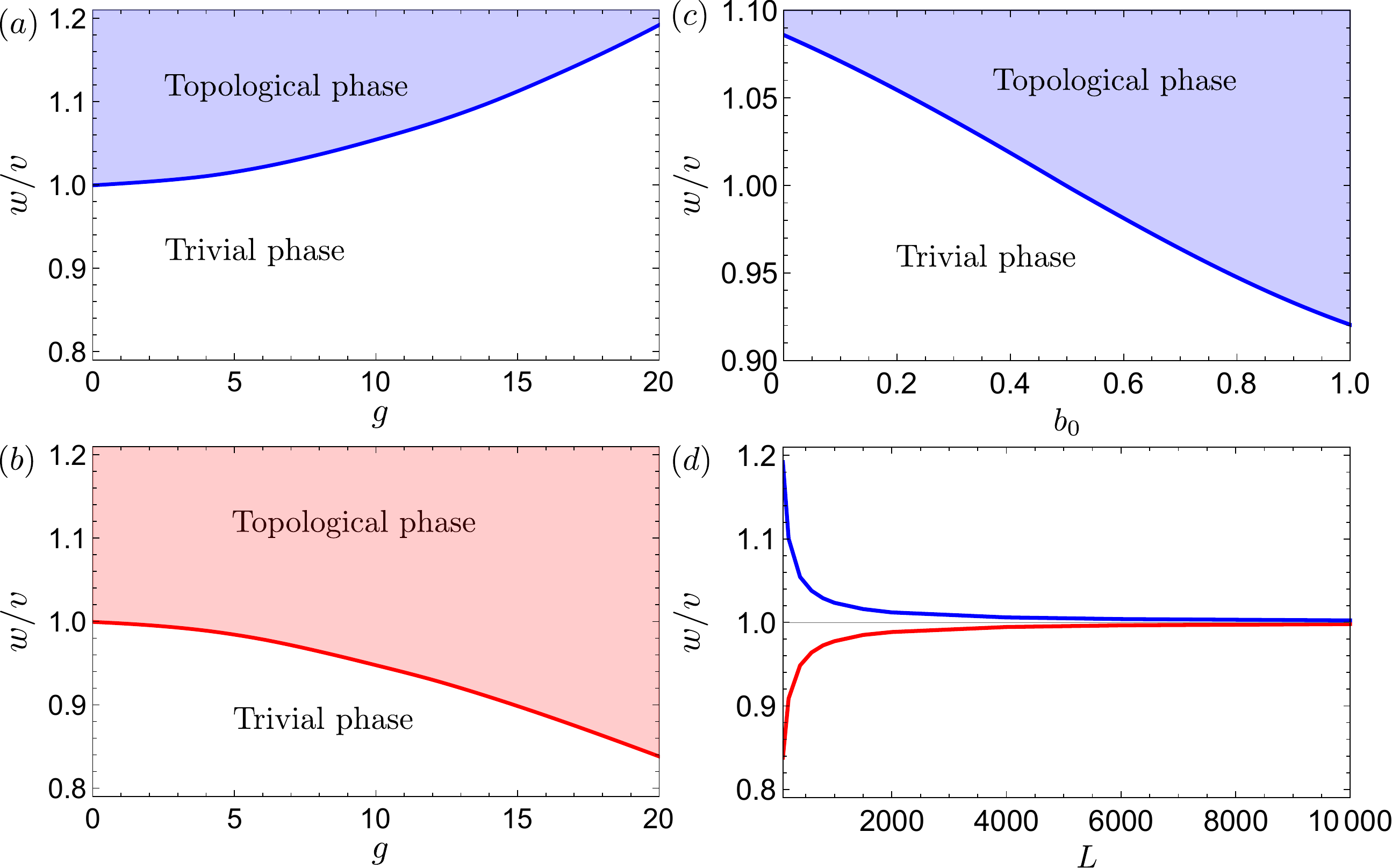}
	\caption{ Topological phase diagram as a function of the light-matter coupling $g$ and intercell hopping $w/v$ for (a) $b_0 = 0.2$ and (b) $b_0 = 0.8$.  Trivial (white area) and topological phases (blue for $b_0 = 0.2$ and red for $b_0 = 0.8$) are separated by the phase boundary (blue and red lines) given by Eq.~\eqref{eq:PhaseTransitionPoint}.
			 Phase boundary is shifted to larger (smaller) values of $w/v$ for $b_0 = 0.2$ ($b_0 = 0.8$). (c) Topological phase diagram as a function of $b_0$ and $w/v$ for a fixed value of the light-matter coupling $g = 10$. The topological phase is entered at smaller (larger) values of $w/v$ for $b_0>0.5$ ($b_0<0.5$).
	(d) Topological phase diagram as a function of $L$ and  $w/v$ for $g = 20$. Phase boundary goes to $w/v = 1$ for large values of $L$ for $b_0 = 0.2$ (blue line) and $b_0 = 0.8$ (red line).
		  Other parameters are fixed as $v=1$,   $a_0 = 1$, $\omega_c = 0.5$, $N_{max} = 10$, and $L=100$ (except panel (d)).
	}
	\label{fig:PhaseDiagram}
\end{figure*}

We find the photonic ground state by solving the associated mean-field photonic Hamiltonian (see Methods). By solving the electronic and photonic mean field Hamiltonian self-consistently, we find that the renormalization factor $\varphi(g,\ell)$ is purely real, a consequence of the fact that the ground-state of the coupled electron-photon does not carry a finite current. The finite-sized spectrum of the effective SSH model is plotted in Fig.~\ref{fig:SpectrumFinite} for different values of the light-matter coupling $g$ and the geometrical factor $b_0$. As expected for $g=0$ we find a topological transition for $w/v>1$, when a pair of exponentially small almost zero modes appear in the middle of the bulk gap. Upon including a finite light-matter coupling we see that the energy spectrum changes depending on the value of the parameter $b_0$. In particular, for $b_0=0.8$ we see that the topological transition is pushed to smaller values of $w/v$, indicating that coupling to the quantum fluctuation of the light field favor the topological phase. On the other hand, for $b_0=0.2$, the situation is reversed and the edge modes appear for larger values of the hopping ratio $w/v$. This different behavior can be understood by looking at the different photonic renormalization to the hopping integrals, as we are going to discuss further below.

{\bf Topological phase diagram.} We now turn to discuss the bulk phase diagram of the coupled electron-photon system and the emergence of topological phase transitions. To this extent we consider a system with periodic boundary conditions and no edges. This allows to write the Hamiltonian in momentum space. Introducing the fermionic field $ \psi^\dag_k = \left(c^\dag_{k,A}, c^\dag_{k,B}\right)$, the Peierls Hamiltonian in the momentum space can be written in compact form as
	\begin{align}
	&H=  \sum_{k} \psi^\dag_k\,\mathcal{H}_k(a,a^{\dagger})\psi_k + \omega_c \left(a^\dag a + \dfrac{1}{2}\right),
	 \label{eq:HamiltonianPeierlsMomentum}
	\end{align}
	where we have introduced an effective single particle electronic Hamiltonian $\mathcal{H}_k(a,a^{\dagger})=\sum_{\alpha}\,d_{k\alpha}(a,a^{\dagger})\sigma_{\alpha}$, with $\sigma_{\alpha=x,y,z}$ Pauli matrices. The structure of this Hamiltonian is encoded in the vector $d_{k\alpha}= \left( d_{kx},d_{ky},d_{kz}\right)$, which depends on the photonic degrees of freedom, the light-matter coupling and the ratio of lattice spacing, and reads
	\begin{align}
	&d_{kx}=v\cos\left[K b_0\right]-w\cos\left[K(1-b_0)\right],\\
	&d_{ky}=-v\sin\left[Kb_0\right]-w\sin\left[K(1-b_0)\right],\\
	&d_{kz}=0,
	\end{align}
	with $K=k+\dfrac{g}{\sqrt{L}}\left(a+a^\dag\right)$ the shifted momentum.

For topological systems such as the SSH model the chiral (or sublattice) symmetry plays a pivotal role, protecting the existence of a transition between two insulating phases with different topological properties. In the language of the pseudo-spin components used in Eq.~(\ref{eq:HamiltonianPeierlsMomentum}) this requires that there is no mass term for the electrons in the $A$ and $B$ sublattices, i.e. $d_{kz}=0$. From the structure of the Hamiltonian in momentum space we see therefore that even in presence of quantum light, $g\neq0$, the chiral symmetry is preserved by the Peierls substitution. This ensures the existence of pairs of eigenstates with opposite energies.

 In the purely electronic model, corresponding to $g=0$, the topological phase transition in the thermodynamic limit would be associated with a closing and reopening of the bulk gap. For a finite-size chain the gap remains finite and the transition point is associated to the minimum energy gap. In presence of light-matter coupling in the collective ultrastrong coupling regime it is important to keep a finite $L$ to obtain non-trivial effects from the photons.  As in the previous section, we employ the mean-field approach to calculate the bulk energy spectrum in the presence of the coupling to photons (see Methods). By solving the electronic and photonic mean-field Hamiltonian self-consistently, we find that the photon renormalization to the electronic hoppings are purely real, and  the electronic bulk energy spectrum reads
\begin{align}
\epsilon_k = \sqrt{\tilde{v}^2+  \tilde{w}^2 - 2 \tilde{v}\tilde{w}  \cos(k)},
\label{eq:epsilonK}
\end{align}
with  $\tilde{v}, \tilde{w}$ defined before Eq.~\eqref{eq:Ema0b0}. The bulk energy gap is given by the value of the dispersion at $k= \pi/L$, $\Delta=\mbox{min}\epsilon_k= \sqrt{\tilde{v}^2+  \tilde{w}^2 - 2 \tilde{v}\tilde{w}  \cos(\pi/L)}$. From this we see that the phase boundary between trivial and topological phases is implicitly given by the equation $\tilde{w}/\tilde{v}=1$, i.e.
\begin{align}
\dfrac{w_c}{v}=\frac{ \varphi(g,b_0)}{\varphi(g,1-b_0)},
\label{eq:PhaseTransitionPoint}
\end{align}
where we emphasize how the renormalization factors depend both on the electronic and photonic parameters, such as light-matter coupling and cavity frequency, as well as the geometric factor $b_0$, through the self-consistent ground state entering Eq.~(\ref{eq:HamiltonianPeierlsMomentum}).

 \begin{figure}[t!]
 	\centering
 	\includegraphics[width=0.9\linewidth]{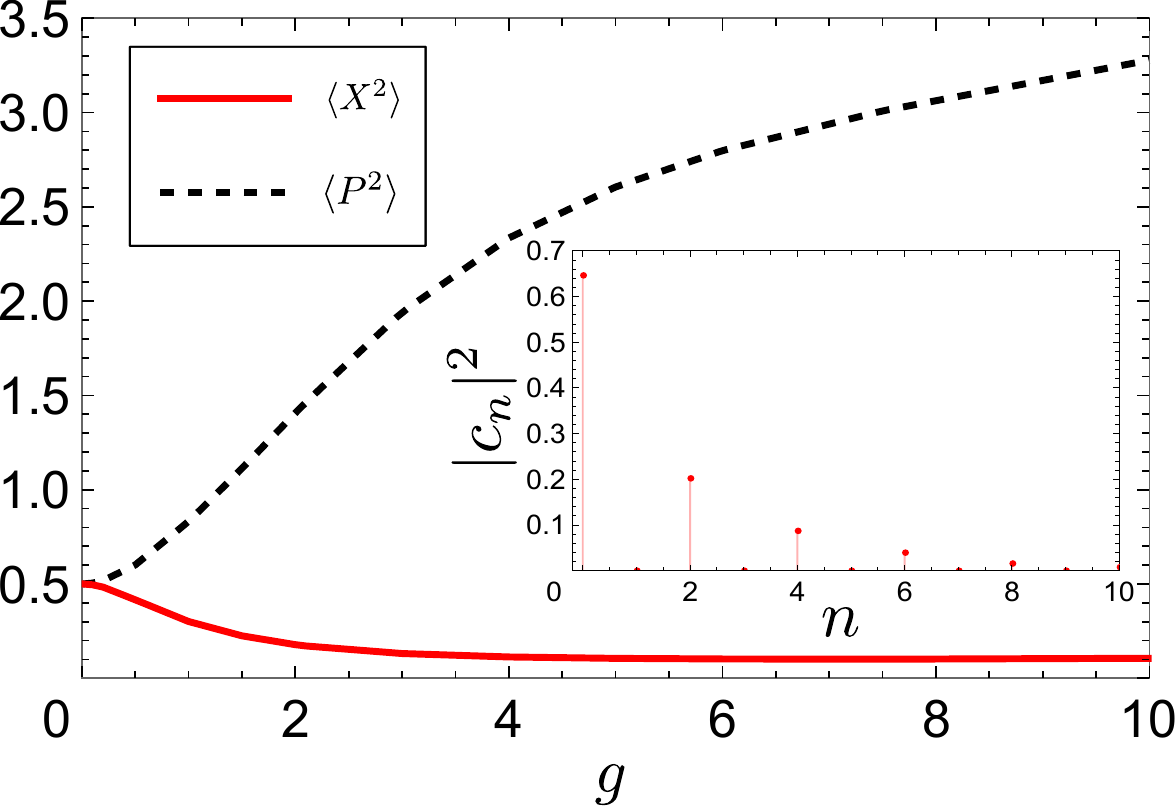}
 	\caption{Expectation values $\langle X^2 \rangle$ (red solid line) and $\langle P^2 \rangle$ (black dashed line) as a function of the light-matter coupling $g$ for $w = v=1$. In the absence of the coupling between the cavity and the SSH chain, the expectation values are equal, $\langle X^2 \rangle = \langle P^2 \rangle = 1/2$. For finite values of $g$, $\langle X^2 \rangle$ decreases and $\langle P^2 \rangle$ increases with increasing $g$, and photonic state becomes squeezed. Inset: Photon probability density $|c_n|^2$ as a function of photon number $n$ for $g = 10$.
 	Other parameters are the same as in Fig.~\ref{fig:SpectrumFinite}.
 	}
 	\label{fig:Squeezing}
 \end{figure}
 
 
Solving the photonic and electronic Hamiltonian self-consistently we find numerically the topological phase diagram that we plot for two values of $b_0$ in Fig.~\ref{fig:PhaseDiagram}  (panel (a) $b_0=0.2$, panel (b) $b_0=0.8$). From this we see, even more clearly, that the light-matter coupling is able to affect the finite-size topological structure of the system. As already observed in the finite-sized edge spectrum, for $b_0=0.2$ we see that coupling to quantum light is energetically detrimental and the topological phase of the SSH model is destroyed.  Strikingly, we see that for $b_0=0.8$ a trivial SSH model can be turned into a topological phase by increasing light-matter interactions. This is one of the main result of this work which opens up the exciting possibility of controlling topology with light. Another interesting aspect is the role played by the geometrical factor $b_0$ in shaping the phase diagram of the system. As we see in Figure~\ref{fig:PhaseDiagram}, panel (c),  also tuning $b_0$ gives rise to different behavior and a possible topological transition. This suggests a comparison with the classical Floquet case~\cite{gomez2013floquet}, that we leave for the discussion.

Finally, we discuss the dependence of the critical hopping strength $w_c$ from the size of the system $L$. As we show in panel (d) of Fig.~\ref{fig:PhaseDiagram} upon increasing $L$ the critical coupling approaches the value $w_c = v$, corresponding to the topological transition in the absence of the cavity ($g = 0$).  This result is consistent with the recent literature stating that single mode cavity are not expected to change bulk properties of the system in the thermodynamic limit~\cite{andolina2019cavity,dmytruk2021gauge,amelio2021optical,eckhardt2021quantum}

 \begin{figure*}[t!]
 	\centering
 	\includegraphics[width=\linewidth]{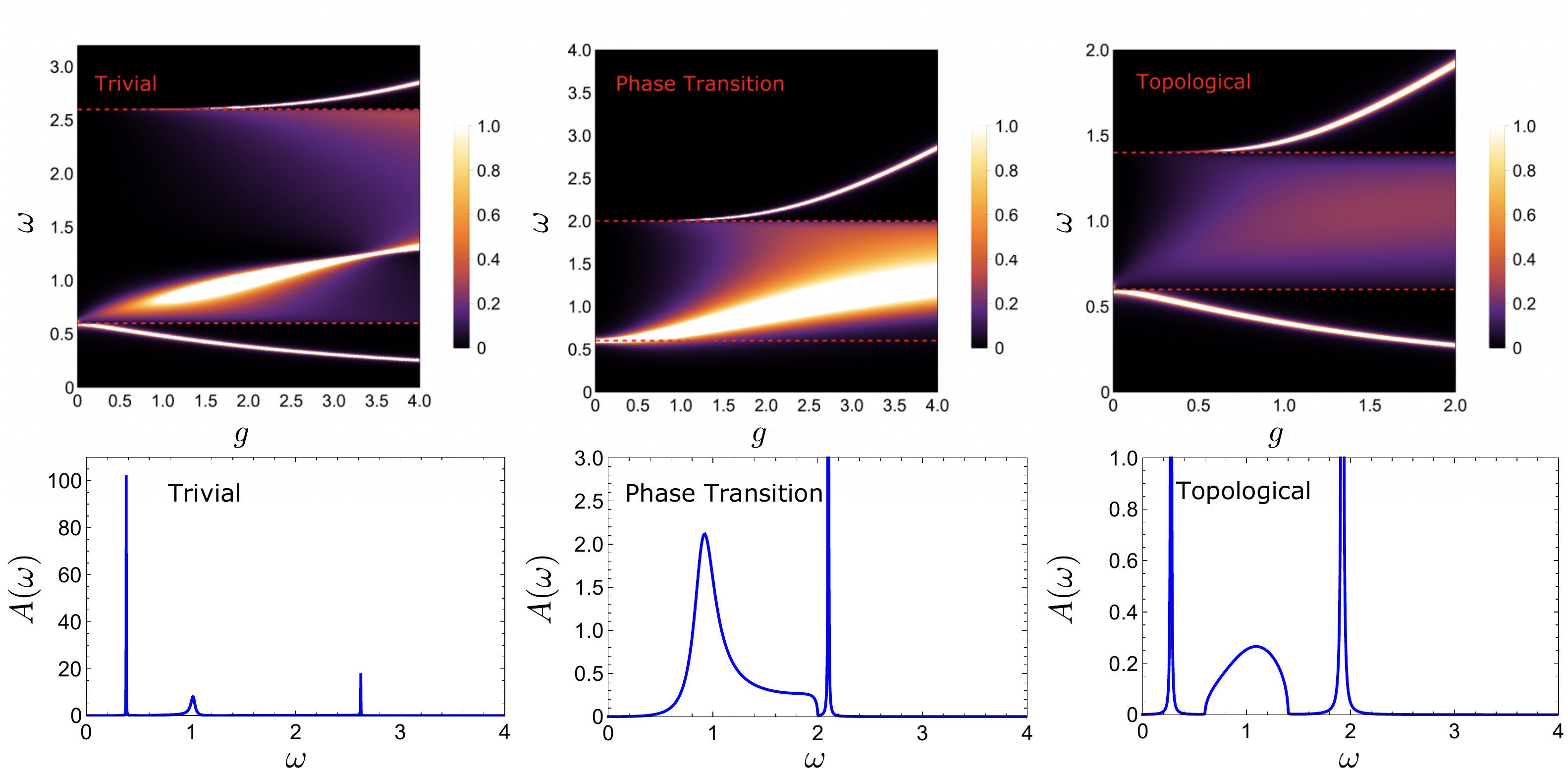}
 	\caption{Spectral function $A(\omega)$ as a function of $g$ and $\omega$.
 		(top panels) Red dashed lines correspond to frequencies $\omega_c$ and $2 E(k = \pi)$.  Left panel: Trivial phase, $v = 0.8$, $w = 0.5$. There are three polariton branches. Middle panel:  The gap at  $k=0$ closes for $v = w = 0.5$ corresponding to the topological phase transition. There are only two polariton branches. Right panel: Topological phase, $v = 0.2$, $w = 0.5$. There are three polariton branches.  The photon frequency $\omega_c$ is chosen such that $\omega_c = 2 E(k=0)$ [for $v \neq w$].  Other parameters are fixed as $\omega_c = 0.6$ and $\eta=10^{-3}$. (bottom panels) Spectral function $A(\omega)$ as a function of $\omega$ for $g = 2$.
 		Left panel: There are three peaks. Middle panel: There are two peaks. Right panel: There are three peaks. The lower polariton branch is absent at the topological phase transition point. 
 	}
 	\label{fig:SpectralFunction}
 \end{figure*}

{\bf Photonic Properties.}  In this section, we focus on the properties of the photonic states coupled to a finite-length SSH chain.  In the basis of the Fock states, the photon ground state can  be written as $|\phi\rangle = \sum_n c_n |n\rangle$, where $n$ denotes the number of photons. The coefficients $c_n$ could be found by calculating the eigenvectors of the photonic Hamiltonian $H_{\rm ph}^{\rm mf}=\langle \psi | H | \psi \rangle $.  We find that the coefficients $c_n$ corresponding to odd number of photons are zero (see inset in Fig.~\ref{fig:Squeezing}).
Next, we introduce the position  $X = \left(a+ a^\dag\right)/\sqrt{2}$ and momentum $P = -i \left(a- a^\dag\right)/\sqrt{2}$ operators.
For a bare photonic Hamiltonian Eq.~\eqref{eq:Hphoton}, the expectation values of the $X^2$ and $P^2$ operators are equal, $\langle X^2 \rangle = \langle P^2 \rangle = 1/2$. However, for a single mode photonic  Hamiltonian coupled to the SSH chain we find that $\langle P^2 \rangle> \langle X^2 \rangle$, therefore,  photonic state becomes squeezed (see Figure~\ref{fig:Squeezing}). We can understood this behavior from the form of the effective photonic Hamiltonian for large $L$ and from the fact that, as we said, there is no coherent displacement induced by the electrons i.e.
\begin{align}
&H_{\rm ph}^{\rm mf} =  \omega_c \left(a^\dag a + \dfrac{1}{2}\right)+r\left(a+a^{\dagger}\right)^2,
\label{eq:HphMF}
\end{align}
where  the squeezing parameter $r$ is controlled by the renormalized kinetic energy of the SSH chain.

{\bf Polariton spectrum as signature of Topological Phase Transition.} Finally, we conclude discussing how the light emitted from the cavity, encoded in the photonic spectral function and measurable through transmission/reflection experiments~\cite{schiro2014tunable,dmytruk2016outofequilibrium,Cottet_2017},  contains crucial signatures of the topological phase transition in the SSH system. The excitations of the SSH chain coupled to a cavity mode are hybrid light-matter polariton quasiparticles whose energies and lifetime can be read out from the photonic spectral function. This strongly depends on the topological properties of the underlying electronic system. Here we show in particular that at the topological transition one polariton mode is pushed to zero frequency and loses spectral weight, thus leaving only two peaks in the spectral function.  Specifically we consider the photon spectral function $A(\omega)=-\dfrac{1}{\pi}\text{Im}\int dt e^{-i\omega t}\left(-i\theta(t)\right)\langle\left[a(t),a^\dag\right]\rangle$, dressed by the electronic system including $1/L$ Gaussian fluctuations~\cite{mazza2019superradiant,dmytruk2021gauge}. This can be written as
	\begin{align}
	A(\omega)= -\dfrac{1}{\pi}\dfrac{\chi''(\omega) (\omega +\omega_c)^2}{(\omega^2-\omega_c^2 -2 \omega_c\chi'(\omega))^2 + (2\omega_c \chi''(\omega) )^2},
	\label{eq:Aomegaapp}
\end{align}
where   $\chi(\omega) =  K(\omega) - \langle J_d \rangle$ is the current-current  correlation function that is a sum of paramagnetic and diamagnetic contributions (see Methods), 
\begin{align}
	&K(\tau - \tau') = -\langle T_c J_p(\tau) J_p(\tau') \rangle,\label{eq:Ktau}\\
&J_p =\dfrac{g}{\sqrt{L}} \sum_k \psi^\dag_k(w \sin(k)\sigma_x - w \cos(k)\sigma_y)\psi_k,\label{eq:ParamagneticCurrent}\\
&J_d = -\dfrac{g^2}{L}\sum_k \psi^\dag_k(w \cos(k)\sigma_x + w \sin(k)\sigma_y)\psi_k.\label{eq:DiamagneticCurrent}
\end{align}
We show the behavior of the photonic spectral function in the top panel of Fig.~\ref{fig:SpectralFunction}. 
In the previous section, we have demonstrated that in the thermodynamic limit the topological phase transition point remains unchanged by the cavity photons even in the (ultra)strong coupling limit and takes place when $v=w$. 
Let us first consider the behavior of the spectral function in the trivial phase ($v>w$).  In the absence of the coupling to the cavity there is only one peak in the spectral function at $\omega_c$. Since we fixed the system parameters such that $\omega_c$  is equal to the band gap $2E(k)$, where $E(k)$ is the electronic bulk energy spectrum given by Eq.~\eqref{eq:epsilonK} in the absence of coupling to cavity, in the SSH model at $k=0$, for finite values of $g$  the peak at $\omega_c$ in the spectral function splits into two. Moreover, there is a third peak appearing in $A(\omega)$ coming from the band gap at $k = \pi$. Therefore, there are three peaks in $A(\omega)$ corresponding to three polariton branches. As we fix the parameters of the SSH chain such that $v=w$, the lowest polariton branch disappears. This is related to closing of the bulk gap at $k = 0$ and is a direct signature of the topological phase transition. Decreasing the intracell hopping $v$ further the SSH model enters the topological phase ($v<w$) and the lowest polariton branch reemerges. To better illustrate the disappearance of the polariton branch at the topological phase transition we plot the cross section of the spectral function for a fixed value of the light-matter coupling constant $g = 2$ as a function of frequency in the lower panel of Fig.~\ref{fig:SpectralFunction}. We note in passing that the width of the second polariton branch is larger in the topological phase.

We provide a simple analytical argument to understand the disappearance of the lowest polariton branch at the phase transition. First, we note that the 
light-matter Hamiltonian in Eq.~\eqref{eq:HamiltonianPeierlsMomentum}, when written in terms of pseudo-spin operators 
$\sigma^k_{\alpha}=\psi^{\dag}_k\sigma_{\alpha}\psi_k$, takes the form of a spin-photon problem often encountered in quantum optics. Here, however, pseudo-spins at different $k$ points are coupled to the same cavity mode.  The lowest excitations of the system can be then obtained from a spin-wave analysis focusing on the $k=0$ sector of the SSH model (see Methods). This allows to write down an effective Hamiltonian in terms of bosonic spin-wave modes $b$ ($b^\dag$) coupled to cavity photons
\begin{align}
&\tilde{H}^{k =0} = \omega_c a^\dag a  + \omega_x b^\dag b - i g w s (a+a^\dag)(b-b^\dag) \notag\\
&- \dfrac{g^2}{2} w s (a+a^\dag)^2   - \dfrac{\omega_x}{2}L,
\label{eq:PeierlsPolaritonH}
\end{align}
where $\omega_x = 2 |v - w|$ and $s = \text{sign}[v - w]$. We find the analytical expression for the polariton energies by using the Bogoliubov-Hopfield transformation~\cite{de2014light}
\begin{align}
\omega_{\pm}^2 = \dfrac{1}{2}\Big(\omega_x^2 + \tilde{\omega}_c^2 \pm\sqrt{(\tilde{\omega}_c^2 - \omega_x^2)^2 + 16 g^2 \omega_c w^2 \omega_x}\Big),
\label{eq:PolaritonBranchGeneral}
\end{align}
where  $\tilde{\omega}_c = \sqrt{\omega_c(\omega_c - 2 w s g^2)}$.  Next, we note that
at $v = w$  the lower polariton branch $\omega_- = 0$ is absent, and the upper polariton frequency is given by $\omega_{+} = \sqrt{\omega_c^2 - 2 \omega_c g^2 w }$. In agreement with the spectral function calculations, the lower polariton branch disappears at the phase transition point providing  a way to probe the phase  transition point in the SSH model via cavity response measurements.

\blank
\textbf{Discussion}\\
In this work, we have shown that coupling an electronic system to the quantum light field of a cavity can be used to control the topological properties of a system and most notably to drive a topological transition with the emergence of edge modes, even in a regime where the uncoupled system would be topologically trivial. We have highlighted these results in the context of paradigmatic model for topological phases of matter, the SSH chain, that we have studied in presence of a finite coupling to a single-mode cavity encoded in the gauge-invariant Peierls phase.

We have shown that quantum light-matter coupling does not spoil the chiral symmetry, which protects the topological phase in the isolated SSH chain and furthermore introduce a non-trivial dependence from the ratio of lattice spacing which can be used as further controlled parameter to control the phase diagram of the system together with the light-matter coupling.  Computing the finite-length energy spectrum as well as the bulk spectrum using a mean-field decoupling of electrons and photons, which treats the Peierls phase non-perturbatively, we have shown that light-matter interaction can turn a trivial insulator into a topological one, depending on the value of $b_0$. Our work therefore provides a simple and paradigmatic example of non-trivial topology induced by quantum light. We emphasize that with respect to the largely explored Floquet engineering schemes, where classical oscillating light fields are used to induce non-trivial topology, our scheme does not suffer from any heating problem, an issue which requires in the Floquet context to assume high-frequency driving.

Finally, we found that in the thermodynamic limit the topological phase transition point $v = w$ in the SSH model is insensitive to cavity photons. In this limit, we evaluated the polariton spectrum of the hybrid electron-photon system that revealed the appearance of three polariton branches for $v \neq w$. Interestingly, we noticed that the lowest polariton branch disappears exactly at the topological phase transition point $v = w$, the result which was further supported by the full analytical solution within the effective spin-wave theory. Therefore, cavity photons provide a way to both control and probe topological properties of the electronic systems. In future works, it would be interesting to consider inhomogeneous electromagnetic field, recently utizlized in the context of superradiance~\cite{nataf2019rashba,guerci2020superradiant,andolina2020theory},  since a momentum-dependent photonic operators might bring a new ingredient to control topological properties of the material.

\blank
\textbf{Methods}\\
\small
\textbf{Gauge-Fixing of Electrons Coupled to Quantum Light.} In this section we derive the Peierls Hamiltonian used in the main text by means of a unitary transformation acting on the electrons. Starting from the operator $\Omega$ defined in Eq.~(\ref{eq:UP}) of the main text we see that the fermionic operators transform as
$$
\Omega^{\dagger} c_{i\mu} \Omega=e^{ie\mathcal{A}R_{i\mu}}\,c_{i\mu}\,.
$$
where $R_{i\mu}=ia_0+\delta_{\mu,B}b_0$.
Applying this unitary operator to the SSH Hamiltonian generates an electron-photon Hamiltonian
$$
H=H_{\rm ph}+\Omega^{\dagger} H_{\rm SSH} \Omega,
$$
 which is equivalent to a renormalization of the hopping integrals according to the Peierls phase, i.e. 
\begin{align}
 c^{\dagger}_{jA}c_{jB}&\rightarrow e^{ig(a+a^{\dagger})b_0}  c^{\dagger}_{jA}c_{jB},\\
 c^{\dagger}_{j+1A}c_{jB}&\rightarrow e^{-ig(a+a^{\dagger})(a_0-b_0)} c^{\dagger}_{j+1A}c_{jB},
\end{align} 
 which reduces to the result in the main text for $a_0=1$ and $g=g/\sqrt{L}$. Furthermore, using the same unitary operator $\Omega$ it is possible to write down the light-matter Hamiltonian in the dipole gauge. This reads
$$
H_D = \Omega H_{\rm ph} \Omega^\dag + H_{\rm SSH},
$$

	\begin{align}
	&H_D =H_{ph} + H_{SSH} +i\dfrac{g \omega_c }{\sqrt{L}} (a-a^\dag) \sum_{j\mu}R_{j\mu}c^{\dagger}_{j\mu}c_{j\mu}
	\notag\\
	 &+ \dfrac{g^2 \omega_c }{L} \sum_{j'j\mu\mu'}R_{j\mu}R_{j'\mu'}c^{\dagger}_{j\mu}c_{j\mu}c^{\dagger}_{j'\mu'}c_{j'\mu'}.
	\end{align}

\textbf{Mean-Field Decoupling Between Electrons and Photons in Finite-Length SSH Chain Coupled to Photons.} We provide details on the solution of the light-matter Hamiltonian $H$ Eq.~\eqref{eq:PeierlsHamiltonian} in mean field that corresponds to neglecting correlations between the cavity modes $|\phi\rangle$  and electrons  $ |\psi\rangle$, 
\begin{align}
|\Psi\rangle = |\psi\rangle |\phi\rangle.
\end{align}
As a result of the mean field decoupling we have to solve an electronic mean field Hamiltonian given by Eq.~\eqref{eq:HelMF}
and  a photonic mean field Hamiltonian that reads
\begin{align}
	&H_{\rm ph}^{\rm mf}= \langle \psi | H | \psi \rangle  =   v \left( \mathcal{C}_{AB}^{jj}  e^{i \frac{g}{\sqrt{L}} b_0 (a+a^\dag)} + \text{h.c.} \right)\notag\\
	&- w\left(\mathcal{C}_{AB}^{j+1,j} e^{-i \frac{g}{\sqrt{L}} (1-b_0) (a+a^\dag)} + \text{h.c.}\right)\notag\\
	&+\omega_c \left(a^\dag a + \dfrac{1}{2}\right) 
	\label{eq:HphMF2}
	\end{align}
where, we introduced two parameters
\begin{align}
	&\mathcal{C}_{AB}^{jj} = \sum_{j=1}^{L}\langle\psi|c^\dag_{jA}c_{jB}|\psi\rangle,\\
	&\mathcal{C}_{AB}^{j+1,j} = \sum_{j=1}^{L-1}\langle\psi|c^\dag_{j+1,A}c_{jB}|\psi\rangle.
	\end{align}

We find the electronic spectrum of the SSH model in presence of coupling to cavity photons by solving Eqs.~\ref{eq:HelMF} and \ref{eq:HphMF2} self-consistently. In particular, we set $g=0$ in $H_{\rm el}^{\rm mf}$ and calculate numerically  $\mathcal{C}_{AB}^\gamma$, where $\gamma = jj $ or $j, j+1$. Then we insert $\mathcal{C}_{AB}^\gamma$ into $H_{\rm ph}^{\rm mf}$ with $g\neq 0$, and evaluate $\tilde{v}$ and $\tilde{w}$. Afterwards, we insert $\tilde{v}$ and $\tilde{w}$ (calculated for $g\neq 0$) into $H_{\rm el}^{\rm mf}$ and evaluate the eigenvalues of the electronic mean-field Hamiltonian. We find that the energy spectrum of the SSH chain is modified in presence of coupling to cavity photons (see Fig.~\ref{fig:SpectrumFinite}).

\textbf{Mean-Field Solution of Momentum Space Hamiltonian.} In order to obtain the topological phase diagram we study the SSH model coupled to cavity in momentum space. Starting from Eq. (\ref{eq:PeierlsHamiltonian}) and performing the Fourier transformation as $c_{j,\mu} = \dfrac{1}{\sqrt{L}}\sum_k e^{i k j} c_{k,\mu}$, we arrive at the  Peierls Hamiltonian in the momentum space given by Eq.~\eqref{eq:HamiltonianPeierlsMomentum}.
Similarly to the finite-length SSH chain, we solve Eq.~\eqref{eq:HamiltonianPeierlsMomentum} within a mean field approximation.
The electronic mean-field Hamiltonian reads
\begin{align}
H_{\rm el}^{\rm mf} = \langle \phi | H | \phi \rangle = \sum_k \Big[ h_x^k \sigma_x^k+ h_y^k\sigma_y^k \Big],
\label{eq:HelKMF}
\end{align}
where we introduced the pseudo-spin operators 
\begin{align}
\sigma^k_{\alpha}=\psi^{\dag}_k\sigma_{\alpha}\psi_k,
\label{eqn:pseudospin}
\end{align}
and 
\begin{align}
		&h_x^k= \langle \phi\vert d_{kx}(a,a^{\dagger}) \vert \phi\rangle,\\
		&h_y^k = \langle \phi\vert d_{ky}(a,a^{\dagger}) \vert \phi\rangle\,.
	\end{align}

At zero temperature, we find for the expectation values over the electronic states 
\begin{align}
&\langle \psi | \sigma_x^k | \psi \rangle = -h_x^k/\epsilon_k,\\ 
&\langle \psi | \sigma_y^k | \psi \rangle = -h_y^k/\epsilon_k,
\end{align}
where we introduced the energy of the SSH model in presence of cavity fields

\begin{align}
\epsilon_k = \sqrt{\left(h_x^k\right)^2 + \left(h_x^y\right)^2}.
\label{eq:epsilonk}
\end{align}
Similarly, the photonic  mean-field Hamiltonian reads
\begin{align}
	&H_{p\rm h}^{\rm mf} = \langle \psi | H | \psi \rangle = \omega_c\left( a^\dag a + \dfrac{1}{2}\right) + \cos\left[\dfrac{g b_0}{\sqrt{L}}(a+a^\dag)\right] \notag\\
	&\times v \sum_k \langle \psi | \sigma_x^k | \psi \rangle +  \sin\left[\dfrac{g b_0}{\sqrt{L}}(a+a^\dag)\right] v \sum_k \langle \psi | \sigma_y^k | \psi \rangle\notag\\
	&-\cos\left[\dfrac{g \left(1-b_0\right)}{\sqrt{L}}(a+a^\dag)\right]  w \sum_k  \Big\{\langle \psi | \sigma_x^k | \psi \rangle \cos(k)\notag \\
	&+ \langle \psi | \sigma_y^k | \psi \rangle \sin(k)\Big\} + \sin\left[\dfrac{g \left(1-b_0\right)}{\sqrt{L}}(a+a^\dag)\right]\notag \\
&\times	w \sum_k  \Big\{  \langle \psi | \sigma_x^k | \psi \rangle \sin (k) - \langle \psi | \sigma_y^k | \psi \rangle \cos(k)\Big\}.
	\label{eq:HphKMF}
	\end{align}

\textbf{Photonic Green's Function - Gaussian Fluctuations}
In this Section, we compute the Gaussian $1/L$ corrections to the photon Green's function~\cite{mazza2019superradiant,dmytruk2021gauge}. For the sake of simplicity we consider the SSH model with $b_0=0$, since the results do not change qualitatively with $b_0$ in the large $L$ limit. The partition function of the SSH model can be written as

\begin{align}
Z =  \int\mathcal{D}\left[\phi,\phi^*,C,C^*\right]e^{-S_{ph}[\phi,\phi^*,C,C^*]  - S_{el}[C,C^*]  - S_{el-ph}[\phi,\phi^*,C,C^*]},
\end{align}
where we separated the different contributions to the total action $S$: $S_{ph}$ describes to the photonic fields, $S_{el}$ corresponds to the purely electronic system, and $S_{el-ph}$ describes to the electron-photon interaction,

\begin{align}
&S_{ph} = -\int_{0}^{\beta}d\tau d\tau ' \phi^*(\tau){d}_0^{-1}(\tau - \tau ')\phi(\tau '),\\
&S_{el} = -\int_{0}^{\beta}d\tau d\tau '\sum_k \left[c^*_{k,A}\partial_\tau c_{k,A} + c^*_{k,B}\partial_\tau c_{k,B} - v \sigma_x^k\right],\\
&S_{el-ph} = - \int_{0}^{\beta}d\tau \cos\left[\dfrac{g}{\sqrt{L}}(\phi(\tau)+\phi^*(\tau))\right]\notag\\
&\times \sum_k \left[w \cos(k)\sigma_x^k + w\sin(k)\sigma_y^k\right]\notag\\
&+\sin\left[\dfrac{g}{\sqrt{L}}(\phi(\tau)+\phi^*(\tau))\right] \sum_k \left[-w\sin(k)\sigma_x^k + w\cos(k)\sigma_y^k\right].\label{eq:elphactionCoulomb}
\end{align}
Here, $\left(\phi^*(\tau), \phi(\tau)\right)$ correspond to the photonic fields, and
\begin{align}
&{d}_0^{-1}(\tau - \tau ') =  -\delta(\tau - \tau ')\left(\partial_\tau + \omega_c\right).
\end{align}
Defining
\begin{align}
Z_0[\phi,\phi^*] =  \int\mathcal{D}\left[C,C^*\right]e^{- S_{el}[C,C^*]  - S_{el-ph}[\phi,\phi^*,C,C^*]},
\end{align}
the partition function can be rewritten as
\begin{align}
Z =  \int\mathcal{D}\left[\phi,\phi^*\right]e^{-S_{ph}[\phi,\phi^*]} Z_0 \equiv  \int\mathcal{D}\left[\phi,\phi^*\right]e^{-S_{eff}},
\end{align}
where
\begin{align}
S_{eff} = S_{ph}-\log Z_0.
\end{align}

 Expanding the action of the SSH model coupled to a single mode cavity up to second order in photonic fields $\phi(\tau)$, we arrive at
 \begin{align}
\tilde{S}_{eff} = \dfrac{1}{2}\int d\tau d\tau' \Phi^\dag(\tau)\left[\mathcal{D}_0^{-1}(\tau - \tau ') - \Pi(\tau-\tau')\right]\Phi(\tau'),
\end{align}
where $\Phi^\dag (\tau)= \left(\phi^*(\tau),\phi(\tau)\right)$, $\mathcal{D}_0^{-1}(\tau - \tau ')$ is the bare photon Green's function, and $\Pi(\tau-\tau')$ is the polarization given by 
\begin{align}
\Pi(\tau - \tau ') = 
\begin{pmatrix}
\dfrac{\delta^2 \log Z_0[\Phi,\Phi^*]}{\delta\phi^*(\tau)\delta \phi(\tau')} & \dfrac{\delta^2 \log Z_0[\Phi,\Phi^*]}{\delta\phi^*(\tau)\delta \phi^*(\tau')}\\
\dfrac{\delta^2 \log Z_0[\Phi,\Phi^*]}{\delta\phi(\tau)\delta \phi(\tau')} & \dfrac{\delta^2 \log Z_0[\Phi,\Phi^*]}{\delta\phi(\tau)\delta \phi^*(\tau')}
\end{pmatrix}\Bigg|_{\Phi(\tau) = \alpha = 0}.
\end{align}
Next, we find that all four components of the polarization are equal and can be written as
\begin{align}
\Pi(\omega) = 
\begin{pmatrix}
1 & 1\\
1& 1
\end{pmatrix}
\chi(\omega),
\end{align}
where $\chi(\omega)$ is the current-current correlation function defined in the main text.

Using Eqs.~\eqref{eq:Ktau}-\eqref{eq:DiamagneticCurrent} and substitution sum with an integral, we find the expressions for the real

	\begin{align}
	&\chi'(\omega) =  \dfrac{g^2}{2\pi} \int_{-\pi}^{\pi} dk \ \dfrac{w (w - v \cos (k))}{E_k}\notag\\ 
	&+\dfrac{2 g^2}{\pi}\mathcal{P}\int_{-\pi}^{\pi} dk \ \left[\dfrac{w (v \cos (k)-w)}{E_k} \right]^2 \dfrac{ E_k}{\omega^2-4E_k^2} 
	\end{align}

and imaginary  parts of the correlation function $\chi(\omega)$

\begin{align}
\chi''(\omega) &= -\dfrac{g^2}{2}\int_{-\pi}^{\pi} dk \ \left[\dfrac{w (v \cos (k)-w)}{E_k} \right]^2\notag\\
&\times \left[\delta\left(\omega-2 E_k\right) - \delta\left(\omega+2 E_k\right)\right].
\end{align}

\textbf{Spin-Wave Theory for Polariton Spectrum.} In this Section, we derive the spin-wave spectrum of the polaritons. To this extent we start from Eq.~\eqref{eq:HamiltonianPeierlsMomentum} that we rewrite in terms of the pseudo-spin operators defined in Eq. (\ref{eqn:pseudospin}). Then,  considering only the momentum $k = 0$ and expanding to second order in $g$, we obtain 
\begin{align}
&\tilde{H}^{k =0} = \omega_c a^\dag a - \dfrac{\omega_x}{2}\tilde{\sigma}_z^{k=0} + \dfrac{g}{\sqrt{L}} w s 
(a+a^\dag) \tilde{\sigma}_y^{k=0}\notag\\
&- \dfrac{g^2}{2L} w s (a+a^\dag)^2 \tilde{\sigma}_z^{k=0} ,
\end{align}
where $\omega_x = 2 |v - w|$ and $s = \text{sign}[v - w]$.
Performing  Holstein - Primakoff transformation~\cite{lerose2019impact,dmytruk2021gauge}
\begin{align}
&\tilde{\sigma}_y^{k = 0} = -i\sqrt{L}(b-b^\dag),\\
&\tilde{\sigma}_z^{k = 0} = L-2 b^\dag b,
\end{align}
we arrive at Eq.~\eqref{eq:PeierlsPolaritonH}.
We find analytical expression for the polariton energies Eq.~\eqref{eq:PolaritonBranchGeneral} by performing the Bogoliubov-Hopfield transformation~\cite{de2014light}.

Since we consider only $k = 0$ mode, it turns out that $\omega_-$ always go to zero for $v \neq w $ at the critical value of the light-matter coupling $g_c = \left|v - w\right|\sqrt{\omega_c}/\sqrt{2 v w |v - w|}$.
We note that the same softening of the lower polariton branch was observed in the case of the excitonic insulator coupled to light using the Peierls substitution~\cite{dmytruk2021gauge} and is absent when all momentum $k$ are taken into account.\\
\textbf{Data availability}\\
Data are available upon reasonable request from the authors.



\begin{thebibliography}{10}
\expandafter\ifx\csname url\endcsname\relax
  \def\url#1{\texttt{#1}}\fi
\expandafter\ifx\csname urlprefix\endcsname\relax\def\urlprefix{URL }\fi
\providecommand{\bibinfo}[2]{#2}
\providecommand{\eprint}[2][]{\url{#2}}

\bibitem{basov2017towards}
\bibinfo{author}{Basov, D.~N.}, \bibinfo{author}{Averitt, R.~D.} \&
  \bibinfo{author}{Hsieh, D.}
\newblock \bibinfo{title}{Towards properties on demand in quantum materials}.
\newblock \emph{\bibinfo{journal}{Nature Materials}}
  \textbf{\bibinfo{volume}{16}}, \bibinfo{pages}{1077--1088}
  (\bibinfo{year}{2017}).
\newblock \urlprefix\url{https://doi.org/10.1038/nmat5017}.

\bibitem{disa2021engineering}
\bibinfo{author}{Disa, A.~S.}, \bibinfo{author}{Nova, T.~F.} \&
  \bibinfo{author}{Cavalleri, A.}
\newblock \bibinfo{title}{Engineering crystal structures with light}.
\newblock \emph{\bibinfo{journal}{Nature Physics}}
  \textbf{\bibinfo{volume}{17}}, \bibinfo{pages}{1087--1092}
  (\bibinfo{year}{2021}).
\newblock \urlprefix\url{https://doi.org/10.1038/s41567-021-01366-1}.

\bibitem{schlawin2021cavity}
\bibinfo{author}{Schlawin, F.}, \bibinfo{author}{Kennes, D.~M.} \&
  \bibinfo{author}{Sentef, M.~A.}
\newblock \bibinfo{title}{Cavity quantum materials} (\bibinfo{year}{2021}).
\newblock \eprint{2112.15018}.

\bibitem{garciavidal2021manipulating}
\bibinfo{author}{Garcia-Vidal, F.~J.}, \bibinfo{author}{Ciuti, C.} \&
  \bibinfo{author}{Ebbesen, T.~W.}
\newblock \bibinfo{title}{Manipulating matter by strong coupling to vacuum
  fields}.
\newblock \emph{\bibinfo{journal}{Science}} \textbf{\bibinfo{volume}{373}},
  \bibinfo{pages}{eabd0336} (\bibinfo{year}{2021}).
\newblock
  \urlprefix\url{https://www.science.org/doi/abs/10.1126/science.abd0336}.
\newblock \eprint{https://www.science.org/doi/pdf/10.1126/science.abd0336}.

\bibitem{valmorra2021vacuum}
\bibinfo{author}{Valmorra, F.} \emph{et~al.}
\newblock \bibinfo{title}{Vacuum-field-induced thz transport gap in a carbon
  nanotube quantum dot}.
\newblock \emph{\bibinfo{journal}{Nature Communications}}
  \textbf{\bibinfo{volume}{12}}, \bibinfo{pages}{5490} (\bibinfo{year}{2021}).
\newblock \urlprefix\url{https://doi.org/10.1038/s41467-021-25733-x}.

\bibitem{hasan2010colloquium}
\bibinfo{author}{Hasan, M.~Z.} \& \bibinfo{author}{Kane, C.~L.}
\newblock \bibinfo{title}{Colloquium: topological insulators}.
\newblock \emph{\bibinfo{journal}{Reviews of Modern Physics}}
  \textbf{\bibinfo{volume}{82}}, \bibinfo{pages}{3045} (\bibinfo{year}{2010}).

\bibitem{qi2011topological}
\bibinfo{author}{Qi, X.-L.} \& \bibinfo{author}{Zhang, S.-C.}
\newblock \bibinfo{title}{Topological insulators and superconductors}.
\newblock \emph{\bibinfo{journal}{Reviews of Modern Physics}}
  \textbf{\bibinfo{volume}{83}}, \bibinfo{pages}{1057} (\bibinfo{year}{2011}).

\bibitem{oka2009photovoltaic}
\bibinfo{author}{Oka, T.} \& \bibinfo{author}{Aoki, H.}
\newblock \bibinfo{title}{Photovoltaic hall effect in graphene}.
\newblock \emph{\bibinfo{journal}{Phys. Rev. B}} \textbf{\bibinfo{volume}{79}},
  \bibinfo{pages}{081406} (\bibinfo{year}{2009}).
\newblock \urlprefix\url{https://link.aps.org/doi/10.1103/PhysRevB.79.081406}.

\bibitem{lindner2011floquet}
\bibinfo{author}{Lindner, N.~H.}, \bibinfo{author}{Refael, G.} \&
  \bibinfo{author}{Galitski, V.}
\newblock \bibinfo{title}{Floquet topological insulator in semiconductor
  quantum wells}.
\newblock \emph{\bibinfo{journal}{Nature Physics}}
  \textbf{\bibinfo{volume}{7}}, \bibinfo{pages}{490--495}
  (\bibinfo{year}{2011}).

\bibitem{wang2013observation}
\bibinfo{author}{Wang, Y.~H.}, \bibinfo{author}{Steinberg, H.},
  \bibinfo{author}{Jarillo-Herrero, P.} \& \bibinfo{author}{Gedik, N.}
\newblock \bibinfo{title}{Observation of floquet-bloch states on the surface of
  a topological insulator}.
\newblock \emph{\bibinfo{journal}{Science}} \textbf{\bibinfo{volume}{342}},
  \bibinfo{pages}{453--457} (\bibinfo{year}{2013}).
\newblock
  \urlprefix\url{https://www.science.org/doi/abs/10.1126/science.1239834}.
\newblock \eprint{https://www.science.org/doi/pdf/10.1126/science.1239834}.

\bibitem{mciver2020light}
\bibinfo{author}{McIver, J.~W.} \emph{et~al.}
\newblock \bibinfo{title}{Light-induced anomalous {H}all effect in graphene}.
\newblock \emph{\bibinfo{journal}{Nature Physics}}
  \textbf{\bibinfo{volume}{16}}, \bibinfo{pages}{38--41}
  (\bibinfo{year}{2020}).

\bibitem{cooper2019topological}
\bibinfo{author}{Cooper, N.~R.}, \bibinfo{author}{Dalibard, J.} \&
  \bibinfo{author}{Spielman, I.~B.}
\newblock \bibinfo{title}{Topological bands for ultracold atoms}.
\newblock \emph{\bibinfo{journal}{Rev. Mod. Phys.}}
  \textbf{\bibinfo{volume}{91}}, \bibinfo{pages}{015005}
  (\bibinfo{year}{2019}).
\newblock
  \urlprefix\url{https://link.aps.org/doi/10.1103/RevModPhys.91.015005}.

\bibitem{karzig2015topological}
\bibinfo{author}{Karzig, T.}, \bibinfo{author}{Bardyn, C.-E.},
  \bibinfo{author}{Lindner, N.~H.} \& \bibinfo{author}{Refael, G.}
\newblock \bibinfo{title}{Topological polaritons}.
\newblock \emph{\bibinfo{journal}{Phys. Rev. X}} \textbf{\bibinfo{volume}{5}},
  \bibinfo{pages}{031001} (\bibinfo{year}{2015}).
\newblock \urlprefix\url{https://link.aps.org/doi/10.1103/PhysRevX.5.031001}.

\bibitem{ohm2015microwave}
\bibinfo{author}{Ohm, C.} \& \bibinfo{author}{Hassler, F.}
\newblock \bibinfo{title}{Microwave readout of majorana qubits}.
\newblock \emph{\bibinfo{journal}{Phys. Rev. B}} \textbf{\bibinfo{volume}{91}},
  \bibinfo{pages}{085406} (\bibinfo{year}{2015}).
\newblock \urlprefix\url{https://link.aps.org/doi/10.1103/PhysRevB.91.085406}.

\bibitem{trif2012resonantly}
\bibinfo{author}{Trif, M.} \& \bibinfo{author}{Tserkovnyak, Y.}
\newblock \bibinfo{title}{Resonantly tunable majorana polariton in a microwave
  cavity}.
\newblock \emph{\bibinfo{journal}{Phys. Rev. Lett.}}
  \textbf{\bibinfo{volume}{109}}, \bibinfo{pages}{257002}
  (\bibinfo{year}{2012}).
\newblock
  \urlprefix\url{https://link.aps.org/doi/10.1103/PhysRevLett.109.257002}.

\bibitem{contamin2021hybrid}
\bibinfo{author}{Contamin, L.~C.}, \bibinfo{author}{Delbecq, M.~R.},
  \bibinfo{author}{Dou{\c c}ot, B.}, \bibinfo{author}{Cottet, A.} \&
  \bibinfo{author}{Kontos, T.}
\newblock \bibinfo{title}{Hybrid light-matter networks of majorana zero modes}.
\newblock \emph{\bibinfo{journal}{npj Quantum Information}}
  \textbf{\bibinfo{volume}{7}}, \bibinfo{pages}{171} (\bibinfo{year}{2021}).
\newblock \urlprefix\url{https://doi.org/10.1038/s41534-021-00508-w}.

\bibitem{wang2019cavity}
\bibinfo{author}{Wang, X.}, \bibinfo{author}{Ronca, E.} \&
  \bibinfo{author}{Sentef, M.~A.}
\newblock \bibinfo{title}{Cavity quantum electrodynamical chern insulator:
  Towards light-induced quantized anomalous hall effect in graphene}.
\newblock \emph{\bibinfo{journal}{Phys. Rev. B}} \textbf{\bibinfo{volume}{99}},
  \bibinfo{pages}{235156} (\bibinfo{year}{2019}).
\newblock \urlprefix\url{https://link.aps.org/doi/10.1103/PhysRevB.99.235156}.

\bibitem{appugliese2022breakdown}
\bibinfo{author}{Appugliese, F.} \emph{et~al.}
\newblock \bibinfo{title}{Breakdown of topological protection by cavity vacuum
  fields in the integer quantum hall effect}.
\newblock \emph{\bibinfo{journal}{Science}} \textbf{\bibinfo{volume}{375}},
  \bibinfo{pages}{1030--1034} (\bibinfo{year}{2022}).
\newblock
  \urlprefix\url{https://www.science.org/doi/abs/10.1126/science.abl5818}.
\newblock \eprint{https://www.science.org/doi/pdf/10.1126/science.abl5818}.

\bibitem{roux2020strongly}
\bibinfo{author}{Roux, K.}, \bibinfo{author}{Konishi, H.},
  \bibinfo{author}{Helson, V.} \& \bibinfo{author}{Brantut, J.-P.}
\newblock \bibinfo{title}{Strongly correlated fermions strongly coupled to
  light}.
\newblock \emph{\bibinfo{journal}{Nature Communications}}
  \textbf{\bibinfo{volume}{11}}, \bibinfo{pages}{2974} (\bibinfo{year}{2020}).
\newblock \urlprefix\url{https://doi.org/10.1038/s41467-020-16767-8}.

\bibitem{mivehvar2017superradiant}
\bibinfo{author}{Mivehvar, F.}, \bibinfo{author}{Ritsch, H.} \&
  \bibinfo{author}{Piazza, F.}
\newblock \bibinfo{title}{Superradiant topological peierls insulator inside an
  optical cavity}.
\newblock \emph{\bibinfo{journal}{Phys. Rev. Lett.}}
  \textbf{\bibinfo{volume}{118}}, \bibinfo{pages}{073602}
  (\bibinfo{year}{2017}).
\newblock
  \urlprefix\url{https://link.aps.org/doi/10.1103/PhysRevLett.118.073602}.

\bibitem{su1979solitons}
\bibinfo{author}{Su, W.}, \bibinfo{author}{Schrieffer, J.} \&
  \bibinfo{author}{Heeger, A.~J.}
\newblock \bibinfo{title}{Solitons in polyacetylene}.
\newblock \emph{\bibinfo{journal}{Physical Review Letters}}
  \textbf{\bibinfo{volume}{42}}, \bibinfo{pages}{1698} (\bibinfo{year}{1979}).

\bibitem{delplace2011zak}
\bibinfo{author}{Delplace, P.}, \bibinfo{author}{Ullmo, D.} \&
  \bibinfo{author}{Montambaux, G.}
\newblock \bibinfo{title}{Zak phase and the existence of edge states in
  graphene}.
\newblock \emph{\bibinfo{journal}{Phys. Rev. B}} \textbf{\bibinfo{volume}{84}},
  \bibinfo{pages}{195452} (\bibinfo{year}{2011}).
\newblock \urlprefix\url{https://link.aps.org/doi/10.1103/PhysRevB.84.195452}.

\bibitem{gomez2013floquet}
\bibinfo{author}{G\'omez-Le\'on, A.} \& \bibinfo{author}{Platero, G.}
\newblock \bibinfo{title}{Floquet-bloch theory and topology in periodically
  driven lattices}.
\newblock \emph{\bibinfo{journal}{Phys. Rev. Lett.}}
  \textbf{\bibinfo{volume}{110}}, \bibinfo{pages}{200403}
  (\bibinfo{year}{2013}).
\newblock
  \urlprefix\url{https://link.aps.org/doi/10.1103/PhysRevLett.110.200403}.

\bibitem{dallago2015floquet}
\bibinfo{author}{Dal~Lago, V.}, \bibinfo{author}{Atala, M.} \&
  \bibinfo{author}{Foa~Torres, L. E.~F.}
\newblock \bibinfo{title}{Floquet topological transitions in a driven
  one-dimensional topological insulator}.
\newblock \emph{\bibinfo{journal}{Phys. Rev. A}} \textbf{\bibinfo{volume}{92}},
  \bibinfo{pages}{023624} (\bibinfo{year}{2015}).
\newblock \urlprefix\url{https://link.aps.org/doi/10.1103/PhysRevA.92.023624}.

\bibitem{goren2018topological}
\bibinfo{author}{Goren, T.}, \bibinfo{author}{Plekhanov, K.},
  \bibinfo{author}{Appas, F.} \& \bibinfo{author}{Le~Hur, K.}
\newblock \bibinfo{title}{Topological {Z}ak phase in strongly coupled {LC}
  circuits}.
\newblock \emph{\bibinfo{journal}{Physical Review B}}
  \textbf{\bibinfo{volume}{97}}, \bibinfo{pages}{041106}
  (\bibinfo{year}{2018}).

\bibitem{downing2019topological}
\bibinfo{author}{Downing, C.~A.}, \bibinfo{author}{Sturges, T.~J.},
  \bibinfo{author}{Weick, G.}, \bibinfo{author}{Stobi\ifmmode~\acute{n}\else
  \'{n}\fi{}ska, M.} \& \bibinfo{author}{Mart\'{\i}n-Moreno, L.}
\newblock \bibinfo{title}{Topological phases of polaritons in a cavity
  waveguide}.
\newblock \emph{\bibinfo{journal}{Phys. Rev. Lett.}}
  \textbf{\bibinfo{volume}{123}}, \bibinfo{pages}{217401}
  (\bibinfo{year}{2019}).
\newblock
  \urlprefix\url{https://link.aps.org/doi/10.1103/PhysRevLett.123.217401}.

\bibitem{nie2021dissipative}
\bibinfo{author}{Nie, W.}, \bibinfo{author}{Antezza, M.}, \bibinfo{author}{Liu,
  Y.-x.} \& \bibinfo{author}{Nori, F.}
\newblock \bibinfo{title}{Dissipative topological phase transition with strong
  system-environment coupling}.
\newblock \emph{\bibinfo{journal}{Phys. Rev. Lett.}}
  \textbf{\bibinfo{volume}{127}}, \bibinfo{pages}{250402}
  (\bibinfo{year}{2021}).

\bibitem{perez2021topology}
\bibinfo{author}{P{\'e}rez-Gonz{\'a}lez, B.},
  \bibinfo{author}{G{\'o}mez-Le{\'o}n, {\'A}.} \& \bibinfo{author}{Platero, G.}
\newblock \bibinfo{title}{Topology detection in cavity qed}.
\newblock \emph{\bibinfo{journal}{arXiv preprint arXiv:2106.08709}}
  (\bibinfo{year}{2021}).

\bibitem{atala2013direct}
\bibinfo{author}{Atala, M.} \emph{et~al.}
\newblock \bibinfo{title}{Direct measurement of the zak phase in topological
  bloch bands}.
\newblock \emph{\bibinfo{journal}{Nature Physics}}
  \textbf{\bibinfo{volume}{9}}, \bibinfo{pages}{795--800}
  (\bibinfo{year}{2013}).
\newblock \urlprefix\url{https://doi.org/10.1038/nphys2790}.

\bibitem{meier2016observation}
\bibinfo{author}{Meier, E.~J.}, \bibinfo{author}{An, F.~A.} \&
  \bibinfo{author}{Gadway, B.}
\newblock \bibinfo{title}{Observation of the topological soliton state in the
  su--schrieffer--heeger model}.
\newblock \emph{\bibinfo{journal}{Nature Communications}}
  \textbf{\bibinfo{volume}{7}}, \bibinfo{pages}{13986} (\bibinfo{year}{2016}).
\newblock \urlprefix\url{https://doi.org/10.1038/ncomms13986}.

\bibitem{deleseleuc2019observation}
\bibinfo{author}{de~Léséleuc, S.} \emph{et~al.}
\newblock \bibinfo{title}{Observation of a symmetry-protected topological phase
  of interacting bosons with rydberg atoms}.
\newblock \emph{\bibinfo{journal}{Science}} \textbf{\bibinfo{volume}{365}},
  \bibinfo{pages}{775--780} (\bibinfo{year}{2019}).
\newblock
  \urlprefix\url{https://www.science.org/doi/abs/10.1126/science.aav9105}.
\newblock \eprint{https://www.science.org/doi/pdf/10.1126/science.aav9105}.

\bibitem{rizzo2018topological}
\bibinfo{author}{Rizzo, D.~J.} \emph{et~al.}
\newblock \bibinfo{title}{Topological band engineering of graphene
  nanoribbons}.
\newblock \emph{\bibinfo{journal}{Nature}} \textbf{\bibinfo{volume}{560}},
  \bibinfo{pages}{204--208} (\bibinfo{year}{2018}).

\bibitem{groning2018engineering}
\bibinfo{author}{Gr{\"o}ning, O.} \emph{et~al.}
\newblock \bibinfo{title}{Engineering of robust topological quantum phases in
  graphene nanoribbons}.
\newblock \emph{\bibinfo{journal}{Nature}} \textbf{\bibinfo{volume}{560}},
  \bibinfo{pages}{209--213} (\bibinfo{year}{2018}).

\bibitem{ozawa2019topological}
\bibinfo{author}{Ozawa, T.} \emph{et~al.}
\newblock \bibinfo{title}{Topological photonics}.
\newblock \emph{\bibinfo{journal}{Rev. Mod. Phys.}}
  \textbf{\bibinfo{volume}{91}}, \bibinfo{pages}{015006}
  (\bibinfo{year}{2019}).
\newblock
  \urlprefix\url{https://link.aps.org/doi/10.1103/RevModPhys.91.015006}.

\bibitem{Solnyshkov:21}
\bibinfo{author}{Solnyshkov, D.~D.} \emph{et~al.}
\newblock \bibinfo{title}{Microcavity polaritons for topological photonics}.
\newblock \emph{\bibinfo{journal}{Opt. Mater. Express}}
  \textbf{\bibinfo{volume}{11}}, \bibinfo{pages}{1119--1142}
  (\bibinfo{year}{2021}).
\newblock
  \urlprefix\url{http://opg.optica.org/ome/abstract.cfm?URI=ome-11-4-1119}.

\bibitem{kim2021quantum}
\bibinfo{author}{Kim, E.} \emph{et~al.}
\newblock \bibinfo{title}{Quantum electrodynamics in a topological waveguide}.
\newblock \emph{\bibinfo{journal}{Phys. Rev. X}} \textbf{\bibinfo{volume}{11}},
  \bibinfo{pages}{011015} (\bibinfo{year}{2021}).
\newblock \urlprefix\url{https://link.aps.org/doi/10.1103/PhysRevX.11.011015}.

\bibitem{huber2016topological}
\bibinfo{author}{Huber, S.~D.}
\newblock \bibinfo{title}{Topological mechanics}.
\newblock \emph{\bibinfo{journal}{Nature Physics}}
  \textbf{\bibinfo{volume}{12}}, \bibinfo{pages}{621--623}
  (\bibinfo{year}{2016}).
\newblock \urlprefix\url{https://doi.org/10.1038/nphys3801}.

\bibitem{li2020electromagnetic}
\bibinfo{author}{Li, J.} \emph{et~al.}
\newblock \bibinfo{title}{Electromagnetic coupling in tight-binding models for
  strongly correlated light and matter}.
\newblock \emph{\bibinfo{journal}{Phys. Rev. B}}
  \textbf{\bibinfo{volume}{101}}, \bibinfo{pages}{205140}
  (\bibinfo{year}{2020}).

\bibitem{sentef2020quantum}
\bibinfo{author}{Sentef, M.~A.}, \bibinfo{author}{Li, J.},
  \bibinfo{author}{K\"unzel, F.} \& \bibinfo{author}{Eckstein, M.}
\newblock \bibinfo{title}{Quantum to classical crossover of floquet engineering
  in correlated quantum systems}.
\newblock \emph{\bibinfo{journal}{Phys. Rev. Research}}
  \textbf{\bibinfo{volume}{2}}, \bibinfo{pages}{033033} (\bibinfo{year}{2020}).
\newblock
  \urlprefix\url{https://link.aps.org/doi/10.1103/PhysRevResearch.2.033033}.

\bibitem{Li2020manipulating}
\bibinfo{author}{Li, J.} \& \bibinfo{author}{Eckstein, M.}
\newblock \bibinfo{title}{Manipulating intertwined orders in solids with
  quantum light}.
\newblock \emph{\bibinfo{journal}{Phys. Rev. Lett.}}
  \textbf{\bibinfo{volume}{125}}, \bibinfo{pages}{217402}
  (\bibinfo{year}{2020}).
\newblock
  \urlprefix\url{https://link.aps.org/doi/10.1103/PhysRevLett.125.217402}.

\bibitem{guerci2020superradiant}
\bibinfo{author}{Guerci, D.}, \bibinfo{author}{Simon, P.} \&
  \bibinfo{author}{Mora, C.}
\newblock \bibinfo{title}{Superradiant phase transition in electronic systems
  and emergent topological phases}.
\newblock \emph{\bibinfo{journal}{Physical Review Letters}}
  \textbf{\bibinfo{volume}{125}}, \bibinfo{pages}{257604}
  (\bibinfo{year}{2020}).

\bibitem{dmytruk2021gauge}
\bibinfo{author}{Dmytruk, O.} \& \bibinfo{author}{Schir{\'o}, M.}
\newblock \bibinfo{title}{Gauge fixing for strongly correlated electrons
  coupled to quantum light}.
\newblock \emph{\bibinfo{journal}{Physical Review B}}
  \textbf{\bibinfo{volume}{103}}, \bibinfo{pages}{075131}
  (\bibinfo{year}{2021}).

\bibitem{ciuti2005quantum}
\bibinfo{author}{Ciuti, C.}, \bibinfo{author}{Bastard, G.} \&
  \bibinfo{author}{Carusotto, I.}
\newblock \bibinfo{title}{Quantum vacuum properties of the intersubband cavity
  polariton field}.
\newblock \emph{\bibinfo{journal}{Physical Review B}}
  \textbf{\bibinfo{volume}{72}}, \bibinfo{pages}{115303}
  (\bibinfo{year}{2005}).

\bibitem{frisk2019ultrastrong}
\bibinfo{author}{Frisk~Kockum, A.}, \bibinfo{author}{Miranowicz, A.},
  \bibinfo{author}{De~Liberato, S.}, \bibinfo{author}{Savasta, S.} \&
  \bibinfo{author}{Nori, F.}
\newblock \bibinfo{title}{Ultrastrong coupling between light and matter}.
\newblock \emph{\bibinfo{journal}{Nature Reviews Physics}}
  \textbf{\bibinfo{volume}{1}}, \bibinfo{pages}{19--40} (\bibinfo{year}{2019}).

\bibitem{pilar2020thermodynamics}
\bibinfo{author}{Pilar, P.}, \bibinfo{author}{De~Bernardis, D.} \&
  \bibinfo{author}{Rabl, P.}
\newblock \bibinfo{title}{Thermodynamics of ultrastrongly coupled light-matter
  systems}.
\newblock \emph{\bibinfo{journal}{Quantum}} \textbf{\bibinfo{volume}{4}},
  \bibinfo{pages}{335} (\bibinfo{year}{2020}).

\bibitem{eckhardt2021quantum}
\bibinfo{author}{Eckhardt, C.~J.} \emph{et~al.}
\newblock \bibinfo{title}{Quantum floquet engineering with an exactly solvable
  tight-binding chain in a cavity} (\bibinfo{year}{2021}).
\newblock \eprint{2107.12236}.

\bibitem{asboth2016short}
\bibinfo{author}{Asb{\'o}th, J.~K.}, \bibinfo{author}{Oroszl{\'a}ny, L.} \&
  \bibinfo{author}{P{\'a}lyi, A.}
\newblock \bibinfo{title}{A short course on topological insulators}.
\newblock \emph{\bibinfo{journal}{Lecture Notes in Physics}}
  \textbf{\bibinfo{volume}{919}}, \bibinfo{pages}{166} (\bibinfo{year}{2016}).

\bibitem{andolina2019cavity}
\bibinfo{author}{Andolina, G.}, \bibinfo{author}{Pellegrino, F.},
  \bibinfo{author}{Giovannetti, V.}, \bibinfo{author}{MacDonald, A.} \&
  \bibinfo{author}{Polini, M.}
\newblock \bibinfo{title}{Cavity quantum electrodynamics of strongly correlated
  electron systems: A no-go theorem for photon condensation}.
\newblock \emph{\bibinfo{journal}{Phys. Rev. B}}
  \textbf{\bibinfo{volume}{100}}, \bibinfo{pages}{121109 (R)}
  (\bibinfo{year}{2019}).

\bibitem{amelio2021optical}
\bibinfo{author}{Amelio, I.}, \bibinfo{author}{Korosec, L.},
  \bibinfo{author}{Carusotto, I.} \& \bibinfo{author}{Mazza, G.}
\newblock \bibinfo{title}{Optical dressing of the electronic response of
  two-dimensional semiconductors in quantum and classical descriptions of
  cavity electrodynamics}.
\newblock \emph{\bibinfo{journal}{Physical Review B}}
  \textbf{\bibinfo{volume}{104}}, \bibinfo{pages}{235120}
  (\bibinfo{year}{2021}).

\bibitem{schiro2014tunable}
\bibinfo{author}{Schir\'o, M.} \& \bibinfo{author}{Le~Hur, K.}
\newblock \bibinfo{title}{Tunable hybrid quantum electrodynamics from nonlinear
  electron transport}.
\newblock \emph{\bibinfo{journal}{Phys. Rev. B}} \textbf{\bibinfo{volume}{89}},
  \bibinfo{pages}{195127} (\bibinfo{year}{2014}).
\newblock \urlprefix\url{https://link.aps.org/doi/10.1103/PhysRevB.89.195127}.

\bibitem{dmytruk2016outofequilibrium}
\bibinfo{author}{Dmytruk, O.}, \bibinfo{author}{Trif, M.},
  \bibinfo{author}{Mora, C.} \& \bibinfo{author}{Simon, P.}
\newblock \bibinfo{title}{Out-of-equilibrium quantum dot coupled to a microwave
  cavity}.
\newblock \emph{\bibinfo{journal}{Phys. Rev. B}} \textbf{\bibinfo{volume}{93}},
  \bibinfo{pages}{075425} (\bibinfo{year}{2016}).
\newblock \urlprefix\url{https://link.aps.org/doi/10.1103/PhysRevB.93.075425}.

\bibitem{Cottet_2017}
\bibinfo{author}{Cottet, A.} \emph{et~al.}
\newblock \bibinfo{title}{Cavity {QED} with hybrid nanocircuits: from
  atomic-like physics to condensed matter phenomena}.
\newblock \emph{\bibinfo{journal}{Journal of Physics: Condensed Matter}}
  \textbf{\bibinfo{volume}{29}}, \bibinfo{pages}{433002}
  (\bibinfo{year}{2017}).
\newblock \urlprefix\url{https://doi.org/10.1088/1361-648x/aa7b4d}.

\bibitem{mazza2019superradiant}
\bibinfo{author}{Mazza, G.} \& \bibinfo{author}{Georges, A.}
\newblock \bibinfo{title}{Superradiant quantum materials}.
\newblock \emph{\bibinfo{journal}{Phys. Rev. Lett.}}
  \textbf{\bibinfo{volume}{122}}, \bibinfo{pages}{017401}
  (\bibinfo{year}{2019}).
\newblock
  \urlprefix\url{https://link.aps.org/doi/10.1103/PhysRevLett.122.017401}.

\bibitem{de2014light}
\bibinfo{author}{De~Liberato, S.}
\newblock \bibinfo{title}{Light-matter decoupling in the deep strong coupling
  regime: {T}he breakdown of the {P}urcell effect}.
\newblock \emph{\bibinfo{journal}{Physical Review Letters}}
  \textbf{\bibinfo{volume}{112}}, \bibinfo{pages}{016401}
  (\bibinfo{year}{2014}).

\bibitem{nataf2019rashba}
\bibinfo{author}{Nataf, P.}, \bibinfo{author}{Champel, T.},
  \bibinfo{author}{Blatter, G.} \& \bibinfo{author}{Basko, D.~M.}
\newblock \bibinfo{title}{Rashba cavity {QED}: a route towards the superradiant
  quantum phase transition}.
\newblock \emph{\bibinfo{journal}{Physical Review Letters}}
  \textbf{\bibinfo{volume}{123}}, \bibinfo{pages}{207402}
  (\bibinfo{year}{2019}).

\bibitem{andolina2020theory}
\bibinfo{author}{Andolina, G.}, \bibinfo{author}{Pellegrino, F.},
  \bibinfo{author}{Giovannetti, V.}, \bibinfo{author}{MacDonald, A.} \&
  \bibinfo{author}{Polini, M.}
\newblock \bibinfo{title}{Theory of photon condensation in a spatially varying
  electromagnetic field}.
\newblock \emph{\bibinfo{journal}{Physical Review B}}
  \textbf{\bibinfo{volume}{102}}, \bibinfo{pages}{125137}
  (\bibinfo{year}{2020}).

\bibitem{lerose2019impact}
\bibinfo{author}{Lerose, A.}, \bibinfo{author}{{\v{Z}}unkovi{\v{c}}, B.},
  \bibinfo{author}{Marino, J.}, \bibinfo{author}{Gambassi, A.} \&
  \bibinfo{author}{Silva, A.}
\newblock \bibinfo{title}{Impact of nonequilibrium fluctuations on prethermal
  dynamical phase transitions in long-range interacting spin chains}.
\newblock \emph{\bibinfo{journal}{Physical Review B}}
  \textbf{\bibinfo{volume}{99}}, \bibinfo{pages}{045128}
  (\bibinfo{year}{2019}).

\end{thebibliography}

\blank
\textbf{Acknowledgments}\\
This project has received funding from the European Union’s Horizon 2020 research and innovation programme under the Marie Skłodowska-Curie Grant Agreement No.~892800 and from the European Research Council (ERC) under the European Union’s Horizon 2020 research and innovation programme (Grant agreement No. 101002955 — CONQUER).

\end{document}